# Dynamic prediction and analysis based on restricted mean survival time in survival analysis with nonproportional hazards [§]

**Zijing Yang[1], Hongji Wu[1], Yawen Hou[2], Hao Yuan[1], Zheng Chen[1*]**

[1] Department of Biostatistics, Southern Medical University, Guangzhou, China

[2] Department of Statistics, Jinan University, Guangzhou, China

[*] Corresponding author: Zheng Chen

**ABSTRACT**   In the process of clinical diagnosis and treatment, the restricted mean survival time (RMST), which reflects the life expectancy of patients up to a specified time, can be used as an appropriate outcome measure. However, the RMST only calculates the mean survival time of patients within a period of time after the start of follow-up and may not accurately portray the change in a patient's life expectancy over time. The life expectancy can be adjusted for the time the patient has already survived and defined as the conditional restricted mean survival time (cRMST). A dynamic RMST model based on the cRMST can be established by incorporating time-dependent covariates and covariates with time-varying effects. We analysed data from a study of primary biliary cirrhosis (PBC) to illustrate the use of the dynamic RMST model. The predictive performance was evaluated using the C-index and the prediction error. The proposed dynamic RMST model, which can explore the dynamic effects of prognostic factors on survival time, has better predictive performance than the RMST model. Three PBC patient examples were used to illustrate how the predicted cRMST changed at different prediction times during follow-up. The use of the dynamic RMST model based on the cRMST allows for optimization of evidence-based decision-making by updating personalized dynamic life expectancy for patients.

## 1. Introduction

Time-to-event outcomes, such as overall survival or progression-free survival, are often used as the primary endpoint for clinical trials in many diseases. In this context, survival curves are estimated by the Kaplan-Meier method, and comparisons are performed by the log-rank test. The hazard ratio (HR) obtained from the Cox proportional hazards (PHs) regression model is used to quantify treatment effects. However, the Cox model must satisfy the PHs assumption that the HR is constant over time, which often fails during long-term follow-up[1,2]. Furthermore, as the ratio of hazard rates (or hazard functions) in the two groups, the HR is difficult to interpret and hard to translate into clinical benefits in terms of a prolonged survival time[3-5]. As an alternative, the restricted mean survival time (RMST) is a good summary of the survival distribution, and the treatment effect can be quantified by the difference in the RMST between two treatment groups[6-8].

Generally, after being diagnosed (such as at the time of diagnosis or after a period of treatment), one of the key questions that is often asked by patients is "How long will I live?". This question can be answered by estimating the mean survival time. For example, Fig. 1 shows the survival curve of patients with primary biliary cirrhosis (PBC) from a clinical trial[9], and the area under the entire curve is their mean survival time. However, the mean survival time cannot be estimated unless follow-up is continued until each subject has experienced the event of interest (or in the presence of censoring, until the survival curve has reached zero)[3]. In Fig. 1, the follow-up time of this trial was actually 14.31 years, and it was impossible to observe survival after the end of follow-up. At this time, the area under the survival curve up to 14.31 years can be calculated, that is, the 14.31-year RMST. It is readily interpretable as the mean survival time or "life expectancy" between the start of follow-up and a specific time point $\tau(\tau = 14.31)$ [7,10,11].

It is worth noting that patients may want to know their prognosis at any time during follow-up, which requires the continuous prediction of life expectancy at a different prediction times, represented by $s$. As shown in Fig. 2, a PBC patient started follow-up at $s = s_0$ and underwent liver transplantation at $s = s_1$. The question "How long will I live?" is equally pressing at $s = s_1$ as it was at the start of follow-up ( $s = s_0$ ). However, the patient's life expectancy may vary at different prediction times. First, in the time between $s_0$ and $s_1$, important events have taken place, such as surgical treatment, that may alter a patient's life expectancy. Second, some variables that have an impact on the outcome may exhibit time-varying effects, resulting in a change in life expectancy as time progresses[12-14]. For instance, due to the possibility of postoperative infection and/or transplant rejection, the life expectancy of this patient will be reduced at $s = s_1$ but then greatly improved at $s = s_2$ if the early postoperative period can be successfully survived. Third, some clinical, biochemical and histological indicators (e.g., coagulation indicators) are often measured in subjects at each follow-up visit; these response data give rise to time-dependent covariates (or longitudinal data). Changes in these indicators will also have an impact on life expectancy.

In view of this, the continually updating life expectancy or mean survival time depending on the prediction time $s$ is defined as the conditional restricted mean survival time (cRMST), represented by $m(s,w)$, that is,

$$m(s, w) = \frac{\int_s^{s+w} S(t)dt}{S(s)},$$

where $S(t)$ denotes the survival function, $s$ is the prediction time (more precisely, the time of the prediction) and $w$ is the time window. For example, $m(0,5)$ represents the life expectancy of the patient in the next 5 years from the start of follow-up, which is equivalent to the 5-year RMST, while $m(3,5)$ means the life expectancy in the next 5 years of a patient who had already survived for 3 years from the start of follow-up. The difference in cRMSTs between groups is represented by $md(s,w)$. This concept of obtaining/updating the life expectancy at different prediction times by considering time-dependent covariates and covariates with time-varying effects is called "dynamic prediction"[15,16].

To illustrate the clinical applicability of dynamic prediction based on the cRMST, we utilized a dataset from a well-known clinical study conducted at Mayo Clinic on the treatment of liver disease[9]. A dynamic prediction model (i.e., dynamic RMST model) was developed by landmarking[15,17,18] to explore the dynamic effects of prognostic factors on survival time. Specific patient examples were used to illustrate how the predicted cRMST changed at different prediction times during follow-up.

# 2.Methods

## *2.1. Data sources*

This example comes from the PBC data collected by the Mayo Clinic from January 1974 to May 1984. Follow-up was extended to April 30, 1988. A total of 312 patients participated in the study, of whom 158 (50.6%) were randomly assigned to receive D-penicillamine and 154 to receive a placebo. Patients had on average 6.23 visits, resulting in a total of 1945 observations. The outcome of this analysis was overall survival, which was calculated in years from the time of referral to death.

There were nine baseline and time-dependent covariates that were included in the dynamic RMST model. Predictors measured at baseline were the drug (D-penicillamine, placebo), sex (female, male) and age (years). The time-dependent covariates were the serum bilirubin value (mg/dl), edema (yes, no), serum albumin value (g/dl), prothrombin time (seconds), histologic stage of disease (I/II, III, IV) and serum glutamic oxaloacetic transaminase (SGOT) level (U/ml).

### 2.2. Statistical analysis

To obtain the dynamic prediction of the 5-year ($w$=5) cRMST, a set of landmark time points ($s_l$) were chosen from the prediction times: in the current model, $s_l (l = 0, 1, ..., 25)$ were selected every 0.2 years from the start of follow-up. For each landmark time point $s_l$, the corresponding landmark dataset $R_l$ was constructed by selecting all patients still alive and undergoing follow-up at $s_l$. Then, $\hat{m}_i(s_l, w)$, the estimator of the cRMST corresponding to each individual $i$ ($i$=1,2,…, $n_l$) in $R_l$, could be calculated (see Supplementary File S1) and used as a dependent variable for a generalized linear model (GLM): $\hat{m}(s_l, w \mid Z(s_l)) = a_l + Z(s_l)^T b_l$. The intercept $a_l$ and coefficients $b_l = (b_{1l}, b_{2l}, ..., b_{9l})$ are the parameters to be estimated, and $Z(s_l) = (Z_1(s_l), Z_2(s_l), ..., Z_9(s_l))$ are the values of the covariates at $s_l$. All these models were then combined into a dynamic RMST model: $\hat{m}(s, w \mid Z(s)) = \alpha(s) + Z(s)^T \beta$, where $\alpha(s) = \alpha_0 + \alpha_1 s + \alpha_2 s^2$ describes how the intercept changes over $s$ and $\beta = (\beta_1, \beta_2, ..., \beta_9)$ are the regression coefficients. The predictions of 5-year life expectancy are possible for any prediction time $s \in [s_0, s_{25}]$.

To test for time-varying covariate effects, interactions between covariates and $s$ were then included in the dynamic RMST model: $\hat{m}(s, w) = \alpha(s) + Z(s)^T \beta(s)$. The parameter function $\beta(s) = (\beta_1(s), \beta_2(s), ..., \beta_9(s))$ is a vector of functions that describes changes in the covariates' effects, and $\beta_j(s) = \beta_{j0} + \beta_{j1} s + \beta_{j1} s^2$ calculates the difference in cRMST resulting from a one-unit increase in the $j$th ($j = 1, 2, ..., 9$) covariates at $s$ (i.e., $md(s, w)$). Initially, all interactions were included in the model, after which the quadratic time interactions were tested and removed if they had no significant effect. The covariates with nonsignificant quadratic time interactions were then tested for linear time interactions. Similarly, only the significant interactions were retained. For numeric stability, the prediction time was standardized using $\bar{s} = s / (s_L - s_0)$. In addition, a "static" RMST regression model[19] with $\tau = 10$ years was established for comparison with the dynamic RMST model in application.

The predictive performances of different models were evaluated by Harrell's C-index[20] and the prediction error[21]. The C-index measures the probability of concordance between the predicted order and the observed order, while the prediction error is the difference between the predicted value and the observed value. A Monte-Carlo cross-validation was used to avoid overoptimism[22]. The data were divided into a training set (a 70% random sample) and a test set (the remaining 30%). Then, the dynamic RMST model was fitted to the training set and used to predict $m(s_l, 5)$ for these patients who were still at risk at $s_l$ in the test set. Performance measures (Harrell's C-index and prediction error) were calculated separately for each $s_l$. The above steps were repeated 200 times to obtain average C-index and prediction error values.

All statistical tests were performed at a two-sided significance level of 0.05, and all analyses were performed using R software (version 3.6.1). The data underlying this article are open source and available in the R package 'JM'. Supplementary File S2 details the R code used to perform the process.

## 3. Results

The number of patients used for this analysis was 312, with a median follow-up of 6.30 years (range: 0.11~14.31 years). During the follow-up period, 140 individuals (55.1%) died. The overall 5-year survival rate was 71.2% (95% CI: 66.3%-76.5%) and 10-year survival

rate was 47.9% (95% CI: 41.3%-55.4%).

### 3.1. Effects of prognostic factors

Table 1 shows the regression coefficients together with the standard error of the covariates included in the dynamic RMST model, and Fig. 3 shows the dynamic coefficients (i.e., difference in 5-year cRMST $md(s,5)$ curves ($w$=5) with 95% confidence intervals). For reference, Table 2 describes the results of the RMST model.

Drug was not statistically significant in the RMST model ($Z$=-0.554, $P$=0.579) in that only the patients' referral or baseline ($s$=0) values for risk factors were used (Table 2). In contrast, in the dynamic RSMT model, patients treated with D-penicillamine had a lower 5-year life expectancy than those taking the placebo. The dynamic coefficient of this covariate can be calculated by the following formula (Table 1):

$$\beta_{j=1}(s) = \beta_{10} + \beta_{11} \times (s/5) = -0.004 - 0.272 \times (s/5), \ s \in [0,5],$$

that is, the $md(s,5)$ between the D-penicillamine group ($Z_1(s)=1$) and the placebo group ($Z_1(s)=0$). The change in $md(s,5)$ over time based on the drug is depicted in Fig. 3A. It can be seen that there was no significant difference (95% CI of $md(0,5)$ contains 0) in 5-year life expectancy between patients treated with different drugs when $s$=0, but the adverse effects of D-penicillamine increased with increasing prediction time $s$ (the upper limit of the 95% CI was less than 0). This may be due to the serious side effects of D-penicillamine, resulting in an increased incidence of adverse events and an increased risk of death[23].

In addition, serum bilirubin was an important prognostic factor in PBC patients, and high serum bilirubin levels negatively affected the life expectancy of patients (Fig. 3B). Female patients had a longer life expectancy than male patients (Fig. 3C), which may be due to the relatively larger proportion of older patients (more than 60 years) among males (42.9%) than females (15.2%). The occurrence of edema decreased the 5-year life expectancy, but the effect decreased with increasing prediction time $s$ (Fig. 3D). High albumin levels appeared to have a protective effect with regard to the 5-year life expectancy, with the $md(s,5)$ increasing from the start of follow-up (Fig. 3E). The prothrombin time also demonstrated a significant time-varying effect on the 5-year cRMST, with the $md(s,5)$ decreasing from the start of follow-up but increasing 3 years after the time of referral (Fig. 3F). As expected, advanced histologic stage (III and IV) were associated with a reduced 5-year life expectancy compared with early stage (I/II). However, the $md(s,5)$ between these groups decreased with increasing prediction time (Fig. 3G1-2). In contrast, the RMST model cannot reflect the time-varying effects of these covariates. Furthermore, age and SGOT level demonstrated time-constant effects on the 5-year cRMST in the dynamic RMST model, although the SGOT level was not statistically significant in the RMST model ($Z$=-1.826, $P$=0.068).

### 3.2. Individual dynamic prediction

In addition to exploring the dynamic effects of covariates on the 5-year cRMST, another important role of the dynamic RMST model is to provide individual dynamic predictions for patients. Three patients were selected from the dataset analyzed herein (see Table 3 for details). Fig. 4 (the solid lines) shows the 5-year cRMSTs of these patients at different prediction times, as derived from the dynamic RMST model. Patient A visited the clinic at the time of referral ($s$=0), that is, no time-dependent covariates were generated. The 5-year life expectancy of this patient remained basically unchanged ($m(0,5)$=2.66, $m(5,5)$=2.93), indicating that her condition was stable (Fig. 4A). Patient B visited the clinic two times ($s$=0 and $s$=0.665). In the time between 0 and 0.665 years, the observed values of some variables changed (i.e., time-dependent covariates were generated), which reduced the 5-year life expectancy of this patient (Fig. 4B). Patient C made annual visits to the Mayo Clinic after her initial referral until her death. She had a total of 6 visits, and the observed values of the time-dependent covariates were different at each visit, which had an impact on the patient's survival (Fig. 4C).

The ($s$+5)-year RMSTs calculated by the RMST model are also shown in Fig. 4 (dashed lines) (e.g., the horizontal axis $s$=0 corresponds to the 5-year RMST from the start of follow-up, and $s$=5 corresponds to the 10-year RMST from the start of follow-up). Since only the information at the start of follow-up ($s$=0) was considered, the trend in the changes in the RMST remains the same under different situations and does not reflect the change in life expectancy over prediction time.

### 3.3. Model assessment

The model assessment measures (Harrell's C-index and prediction error) were obtained by the 5-year ($w$=5) cRMST in the dynamic RMST model from each landmark time point $s_l$ (solid lines in Fig. 5). Meanwhile, the predictive performances of these corresponding RMST models ($\tau = s_l + w$) were also evaluated (dashed lines in Fig. 5). Compared with the RMST models, the advantages of the dynamic RMST model (a higher C-index and a lower prediction error) are more obvious with increasing prediction time $s$.

### 4. Discussion

Survival prediction is an indispensable integral part of current clinical practice; it can help determine optimal treatment strategies for individual patients and avoid overtreatment and the associated waste of medical resources. Compared with the survival rate, hazard rate and so on, the RMST is directly based on the concept of time, reflecting the life expectancy of patients up to the specified time, and therefore is a more appropriate evaluation measure[24]. In addition, the difference in the RMSTs measures the impact of different treatments on survival and can be a practical and useful alternative to the HR[7,25].

However, the RMST only calculates the mean survival time of patients within a period of time after the start of follow-up ($s = 0$) and may not accurately portray the change in a patient's life expectancy over time. Taking the perspective of a patient who has already survived a number of years, the cRMST, which is the measure proposed in this article that is based on the RMST, provides more relevant information by adjusting the life expectancy for the time the patient has already survived. In a sense, cRMST can also be understood as the restricted mean residual life[26].

Generally, after considering the concept of condition, the estimated value of measures (such as conditional survival and cRMST) will increase as the number of years survived increases. This relationship is usually even more obvious in patients with advanced-stage disease[27,28]. For example, in this dataset, the 5-year cRMST of patients with histologic stage IV disease was 3.44 years at the time of referral (i.e. $m(0,5)=3.44$). If the patient was still alive at 3 or even 5 years after referral, the 5-year cRMST would change to 3.93 ($m(3,5)=3.93$) years and 4.07 ($m(5,5)=4.07$) years. This means an approximately 0.63-year increase in the 5-year life expectancy of patients who have been followed up for 5 years compared with those who have just been referred. This relationship actually reflects a natural selection effect[29]: due to the existence of individual differences in prognosis, patients with a high risk of death are very likely to experience their endpoint events in the initial years after the start of follow-up. Over time, as these patients expire, the surviving population becomes "healthier" and has a longer life expectancy. The concept of the cRMST is a way to quantify this phenomenon and make it easier for clinicians and patients to comprehend. Therefore, for patients who have been alive for a period of time, the cRMST provides valuable and relevant information on how their life expectancy develops over time. This knowledge can help motivate a patient to continue treatment, improve compliance, and ultimately improve survival.

In this paper, based on the cRMST, a dynamic RMST model was established by incorporating time-dependent covariates and allowing for time-varying effects, enabling the updating of the 5-year cRMST for PBC patients at any prediction time $s \in [s_0, s_{25}]$. The continuous prediction of the cRMST during follow-up allows for the optimization of evidence-based decision-making and may improve the personalization of the treatment options for patients with progressive disease. In addition, compared with the RMST models that only use the patients' baseline ($s$=0) risk factors, the dynamic RMST model has better predictive performance, as assessed by the C-index and prediction error.

However, we must pay attention to several points when applying the dynamic RMST model. First, the time window $w$ used depends on the severity of the disease. For severe diseases, $w$=1 or $w$=2 years is relevant, while for milder diseases with longer follow-up times, such as cirrhosis, $w$=5 or even $w$=10 years is reasonable. Second, the selection of landmark time points $s_l$, which implicitly defines the weighting of the prediction time, is independent of the actual event time. The simplest approach is taking these points equidistantly in the selected interval ($s \in [s_0, s_{25}]$). A number of time points between 20 and 100 will be sufficient[17]. Finally, the functional form, such as the quadratic functions used in this study, of $\alpha(s)$ (how to interpret changes over $s$) and $\beta(s)$ (time-varying covariate effect) should be prespecified in practice.

In summary, predicting patient survival is a complex decision-making process involving the patient's own factors, the disease itself, treatment programs, living environment and other factors. Although prediction models can help clinicians improve the accuracy of prediction, the prediction results cannot be blindly accepted. As Lau[30] said, "*every patient is unique, one can only observe and not determine the final journey*".

**Ethical approval:** The data in this study is available in R package JM. No new clinical data was gathered or used. There is thus no need for an ethical approval.

**Funding:** This research was supported by the National Natural Science Foundation of China [grant numbers 81673268, 81903411]; Natural Science Foundation of Guangdong Province [grant number 2018A030313849] and the Guangdong Basic and Applied Basic Research Foundation [grant number 2019A1515011506].

**Conflict of Interest Statement:** The authors declare no competing or financial interest in this work.

**Table 1.** The results of the dynamic RMST model (*w*=5 years)

| Variable | No. ( Deaths ) | Time function[a] | Coefficient | SE | *P* |
|---|---|---|---|---|---|
| **(Intercept)** | | 1 | 7.772 | 0.541 | <0.001 |
| | | $s/5$ | -13.624 | 2.047 | <0.001 |
| | | $(s/5)^2$ | 11.783 | 2.094 | <0.001 |
| **Drug** (ref: placebo) | 154(69) | | | | |
| D-penicillamine | 158(71) | 1 | -0.004 | 0.047 | 0.925 |
| | | $s/5$ | -0.272 | 0.094 | 0.004 |
| **Sex** (ref: male) | 36(26) | | | | |
| Female | 276(114) | 1 | 0.221 | 0.124 | 0.075 |
| | | $s/5$ | 1.738 | 0.689 | 0.012 |
| | | $(s/5)^2$ | -2.353 | 0.726 | 0.001 |
| **SerBilir** (per 1 mg/dl) | 312(140) | 1 | -0.118 | 0.010 | <0.001 |
| | | $s/5$ | -0.147 | 0.051 | 0.004 |
| | | $(s/5)^2$ | 0.164 | 0.055 | 0.003 |
| **Edema** (ref: no) | 247(96) | | | | |
| Yes | 65(44) | 1 | -0.566 | 0.079 | <0.001 |
| | | $s/5$ | 0.311 | 0.150 | 0.038 |
| **Albumin** (per 1 gm/dl) | 312(140) | 1 | 0.278 | 0.074 | <0.001 |
| | | $s/5$ | 0.482 | 0.152 | 0.001 |
| **Prothrombin**(per 1 second) | 312(140) | 1 | -0.261 | 0.043 | <0.001 |
| | | $s/5$ | 0.997 | 0.182 | <0.001 |
| | | $(s/5)^2$ | -0.945 | 0.188 | <0.001 |
| **Histologic** (ref: Ⅰ/Ⅱ) | 83(22) | | | | |
| Ⅲ | 120(48) | 1 | -0.145 | 0.046 | 0.001 |
| | | $s/5$ | 0.280 | 0.104 | 0.007 |
| Ⅳ | 109(70) | 1 | -0.567 | 0.060 | <0.001 |
| | | $s/5$ | 0.516 | 0.118 | <0.001 |
| **Age** (per 1 year) | 312(140) | 1 | -0.021 | 0.002 | <0.001 |
| **SGOT** (per 10 U/ml) | 312(140) | 1 | -0.011 | 0.002 | <0.001 |

Abbreviations: RMST: restricted mean survival time; No: number; SE: standard error; SGOT: serum glutamic oxaloacetic transaminase.

[a]: The effects for covariates are calculated by the following formula:

$$\beta_j(s) = \beta_{j0} + \beta_{j1}(s/5) + \beta_{j2}(s/5)^2,$$

and the intercept for this model is calculated as $\alpha(s) = \alpha_0 + \alpha_1(s/5) + \alpha_2(s/5)^2$.

**Table 2.** The results of the RMST model ( $\tau = 10$ years)

| Variable | Coefficient | 95% CI | Z | P |
|---|---|---|---|---|
| **(Intercept)** | 13.888 | (8.625, 19.151) | 5.172 | <0.001 |
| **Drug** (ref: placebo) | | | | |
| D-penicillamine | -0.153 | (-0.693, 0.387) | -0.554 | 0.579 |
| **Sex** (ref: male) | | | | |
| Female | 1.041 | (0.014, 2.068) | 1.987 | 0.047 |
| **SerBilir** (per 1 mg/dl) | -0.244 | (-0.312, -0.175) | -6.977 | <0.001 |
| **Edema** (ref: no) | | | | |
| Yes | -0.659 | (-1.443, 0.126) | -1.646 | 0.100 |
| **Albumin** (per 1 gm/dl) | 1.432 | (0.685, 2.179) | 3.758 | <0.001 |
| **Prothrombin** (per 1 second) | -0.707 | (-1.057, -0.356) | -3.947 | <0.001 |
| **Histologic** (ref: Ⅰ/Ⅱ) | | | | |
| Ⅲ | -0.761 | (-1.354, -0.168) | -2.513 | 0.012 |
| Ⅳ | -1.409 | (-2.141, -0.677) | -3.771 | <0.001 |
| **Age** (per 1 year) | -0.051 | (-0.078, -0.023) | -3.610 | <0.001 |
| **SGOT** (per 10 U/ml) | -0.056 | (-0.116, 0.004) | -1.826 | 0.068 |

Abbreviations: RMST: restricted mean survival time; CI: confidence interval; SGOT:

serum glutamic oxaloacetic transaminase.

**Table 3**. The definition of example patients

| Patient | Variables | | | | | | | | | |
|---|---|---|---|---|---|---|---|---|---|---|
| | Time[a] | Drug | SerBilir | Sex | Edema | Albumin | Prothrombin | Histologic | Age | SGOT |
| A | 0.000 | D-penicil | 1.4 | female | Yes | 3.13 | 12.2 | Ⅳ | 77 | 86.8 |
| B | 0.000 | D-penicil | 2.4 | male | No | 3.83 | 10.3 | Ⅲ | 35 | 127.0 |
| B | 0.665 | D-penicil | 3.0 | male | No | 3.75 | 10.5 | Ⅳ | 35 | 161.0 |
| C | 0.000 | placebo | 5.2 | female | No | 3.68 | 9.9 | Ⅲ | 52 | 165.9 |
| C | 0.545 | placebo | 6.6 | female | No | 2.87 | 11.4 | Ⅲ | 52 | 196.9 |
| C | 1.035 | placebo | 5.8 | female | No | 2.94 | 10.4 | Ⅳ | 52 | 210.8 |
| C | 2.029 | placebo | 6.0 | female | Yes | 2.60 | 11.6 | Ⅳ | 52 | 207.7 |
| C | 2.984 | placebo | 9.0 | female | Yes | 2.54 | 10.4 | Ⅳ | 52 | 241.0 |
| C | 3.926 | placebo | 16.2 | female | Yes | 1.81 | 12.5 | Ⅳ | 52 | 241.0 |

Abbreviations: SGOT: serum glutamic oxaloacetic transaminase.

[a]: The observation time.

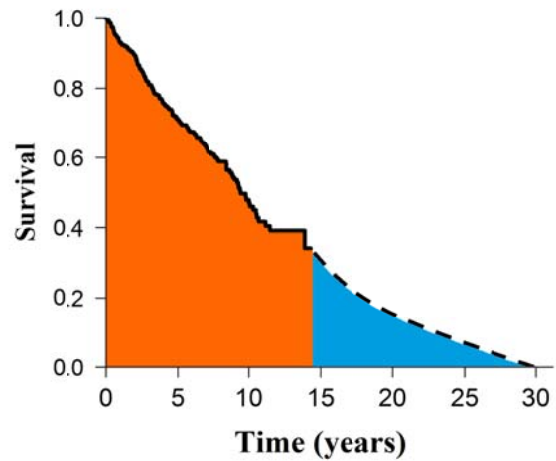

**Figure 1**. Schematic diagram of restricted mean survival time.

The dashed part of the survival curve is extrapolated from the actual observed data (the solid part). The area under the survival curve represents the (unknown) mean survival time, in which the orange area is the 14.31-year RMST ($\tau=14.31$) and the blue area demonstrates the unobserved survival.

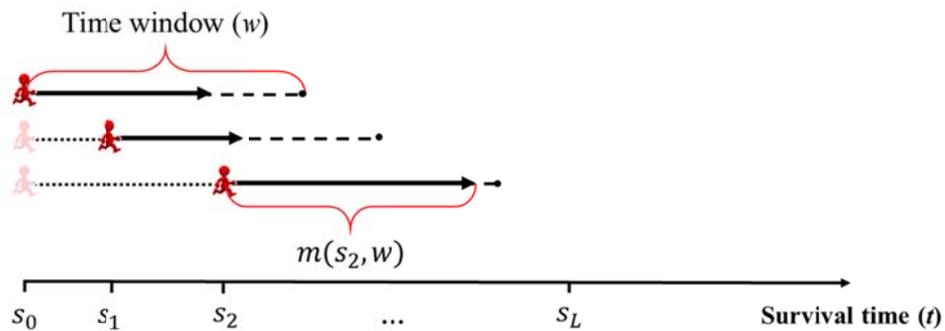

**Figure 2**. Schematic diagram of the dynamic prediction process.

Each line corresponds to the patient's different prediction situations. Here, $w$ is the time window, and $s_l (l = 0, 1, 2, ..., L)$ represents a series of landmark time points, where $s_0$ indicates the start of follow-up. The dotted line (……) means the time the patient has survived, and the solid line is the life expectancy of $w$ years $m(s_l, w)$.

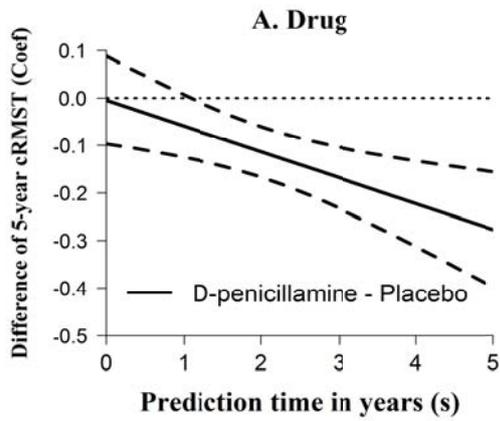

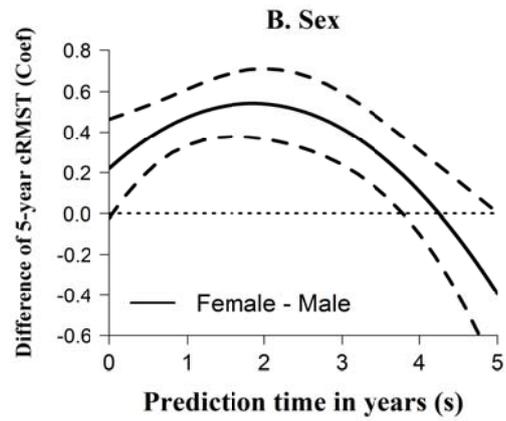

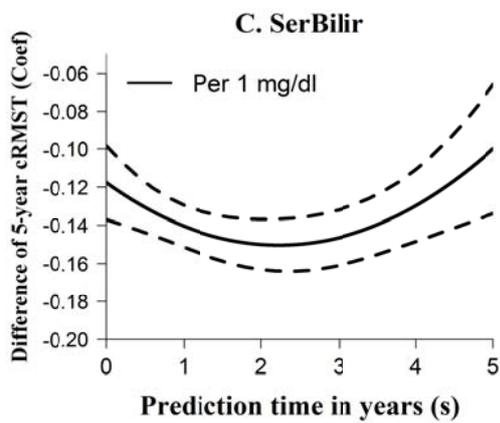

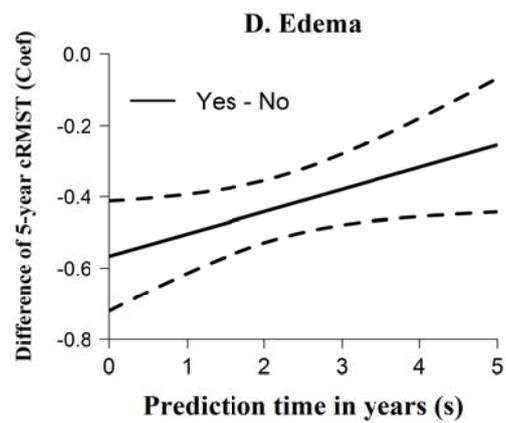

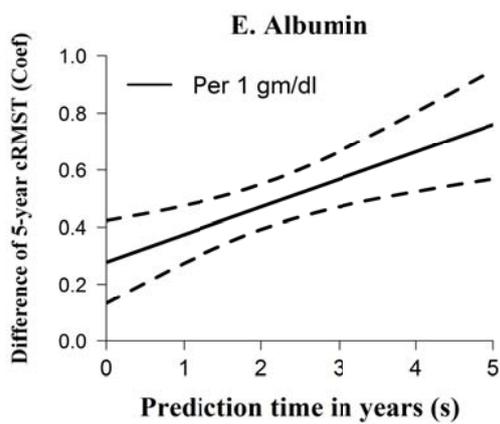

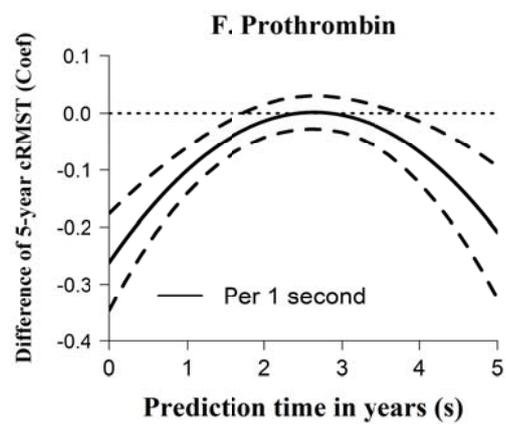

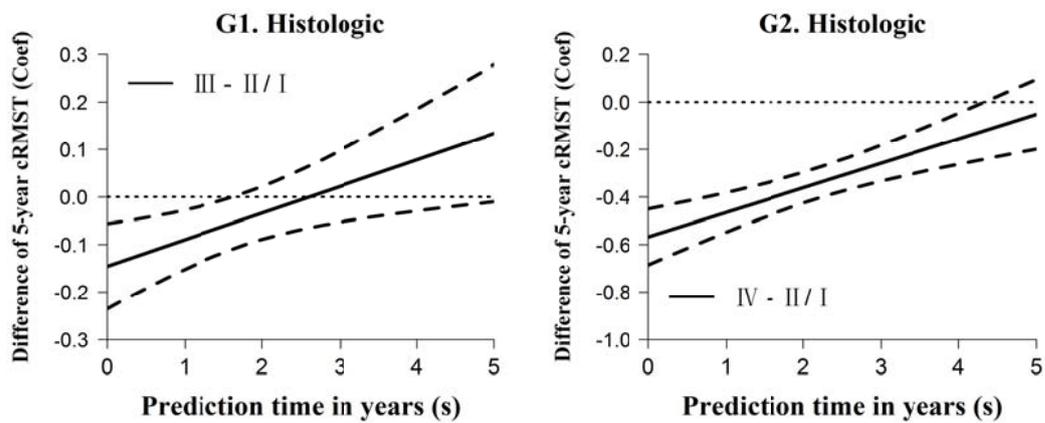

**Figure 3**. Differences of 5-year cRMST (dynamic coefficients $\beta_j(s)$) with 95%

confidence intervals in the dynamic RMST model (*w*=5 years).

Abbreviations: cRMST: conditional restricted mean survival time; Coef: coefficients.

The letters A-G represent different variables, and the numbers 1-2 represent the different levels of Histologic variable. The solid line represents the dynamic coefficients $\beta_j(s)$, that is, the difference in cRMST resulting from a one-unit increase in the *j*th covariates at *s* (*md*(*s*,*w*)). The dashed line (---) represents the 95% confidence intervals.

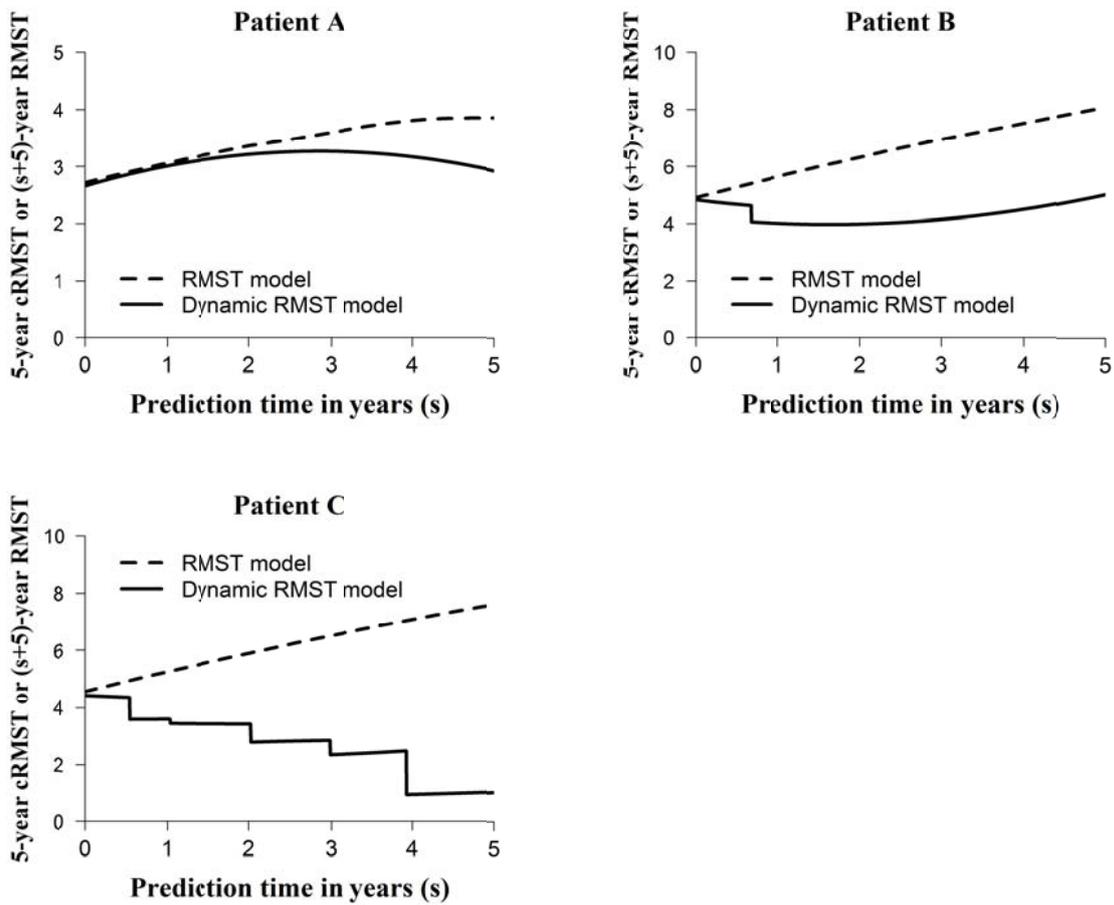

**Figure 4.** Individual predictions with the dynamic RMST model (*w*=5 years) and the

RMST model ($\tau = s + 5$ years).

Abbreviations: cRMST: conditional restricted mean survival time; RMST: restricted mean survival time.

Predictions are shown for three example patients (described in Table 3). The solid lines represent the 5-year cRMST from *s* to *s*+5 years predicted by the dynamic RMST model. The dashed line represents the (*s*+5)-year RMST calculated by the RMST model. For example, the horizontal axis *s*=0 corresponds to the 5-year RMST from the start of follow-up, and *s*=5 corresponds to the 10-year RMST from the start of follow-up.

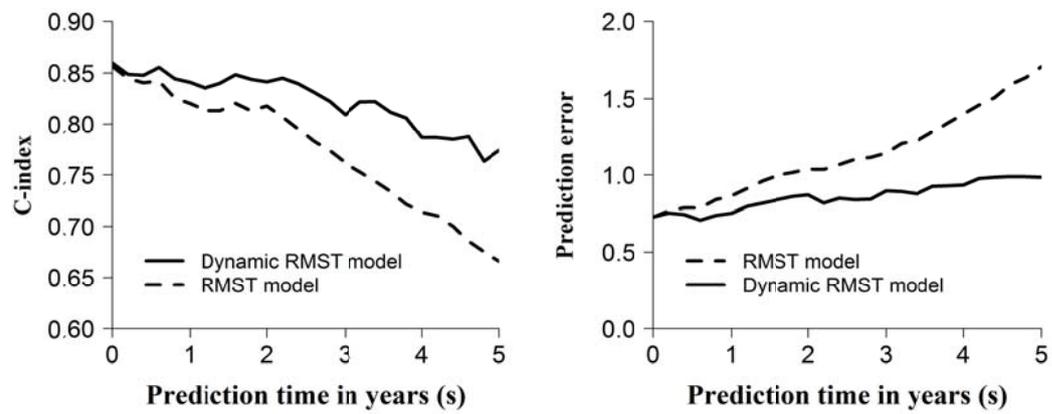

**Figure 5.** Landmark time-specific C-indexes and prediction errors.

Abbreviations: RMST: restricted mean survival time.

A higher C-index indicates better performing model; a lower prediction error indicates better performing model.